\input{aipcheck}

\documentclass[
    ,final            % use final for the camera ready runs
  ]
  {aipproc}

\layoutstyle{6x9}

\begin{document}
\newcommand{\be}{\begin{equation}}
\newcommand{\ee}{\end{equation}}
\newcommand{\bea}{\begin{eqnarray}}
\newcommand{\eea}{\end{eqnarray}}
\newcommand{\bT}{{${\bf b}_\perp$}}
\newcommand{\xT}{{${\bf x}_\perp$}}
\newcommand{\DT}{{${\bf \Delta}_\perp$}}

\title{Generalized Parton Distributions, Deeply Virtual Compton
Scattering and TMDs}

%<Replace this text with PACS numbers; choose from this list:
%                \texttt{http://www.aip..org/pacs/index.html}>}
%      {<Enter Keywords here>}

\author{Matthias Burkardt}{
  address={Department of Physics,
New Mexico State University, Las Cruces, NM 88003.}
}

\begin{abstract}
Parton distributions in impact parameter space, which
are obtained by Fourier transforming GPDs, exhibit
a significant deviation from axial symmetry when 
target and/or quark are transversely polarized. In
combination with the final state interactions, this
transverse deformation provides a natural mechanism for
naive-T odd transverse single-spin asymmetries in
semi-inclusive DIS. 
The deformation of PDFs in impact parameter space can also be related
to the transverse force acting on the active quark
in polarized DIS at higher twist. 
\end{abstract}

\maketitle

%%%%%%%%%%%%%%%%%%%%%%%%%%%%%%%%%%%%%%%%%%%%
%% MAINMATTER
%%%%%%%%%%%%%%%%%%%%%%%%%%%%%%%%%%%%%%%%%%%%

\section{Distribution of Quarks in the Transverse Plane}

\begin{figure}
\unitlength1.cm
\begin{picture}(10,5)(2.7,12.7)
\includegraphics{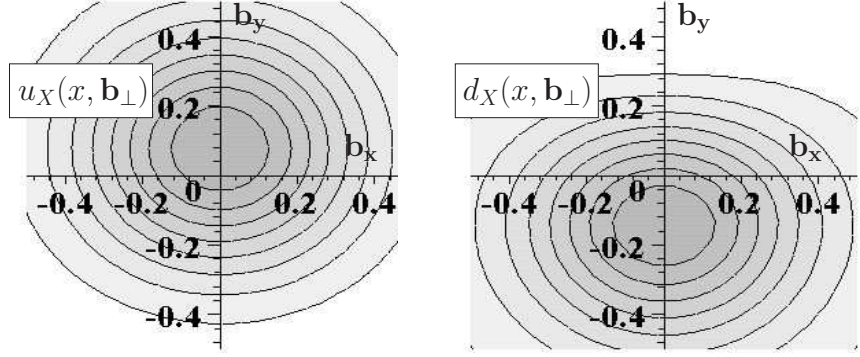}
\end{picture}
\caption{Distribution of the $j^+$ density for
$u$ and $d$ quarks in the
$\perp$ plane ($x=0.3$ is fixed) for a proton that is 
polarized
in the $x$ direction in the model from
Ref. \cite{IJMPA}.
For other values of $x$ the distortion looks similar.
}
\label{fig:distort}
\end{figure}

In the case of transversely polarized quarks and/or nucleons, parton
distributions in impact parameter space show a significant 
transverse deformation. In the case of unpolarized quarks in a nucleon
polarized in the $+\hat{x}$ direction, this deformation is described
by the $\perp$ gradient of the Fourier transform of the GPD $E^q$ 
\cite{IJMPA}
\be
q_{q/p\uparrow}(x,{\bf b}_\perp) = \int \frac{d^2{\bf x}_\perp}{(2\pi)^2}
e^{-i{\bf b}_\perp \cdot { \Delta}_\perp} 
H^q(x,0,-\Delta_\perp^2) - 
\frac{1}{2M} \partial_y \int \frac{d^2{\bf x}_\perp}{(2\pi)^2}
e^{-i{\bf b}_\perp \cdot {\Delta}_\perp} 
E^q(x,0,-{ \Delta}_\perp^2)
\ee
for quarks of flavor $q$. Since $E^q(x,0,t)$ also arises in the 
decomposition of the Pauli form factor $F_2^q=\int_{-1}^1 dx
E^q(x,0,t)$ for quarks with flavor $q$ (here it is always understood
that charge factors have been taken out) w.r.t. $x$, this allows 
to relate the $\perp$ flavor dipole moment to the contribution from
quarks with flavor $q$ to the nucleon anomalous magnetic moment
(here it is always understood
that charge factors have been taken out)
\be
d^q \equiv \int d^2{\bf b}_\perp q_{+\hat{x}}(x,{\bf b}_\perp)
b_y = \frac{1}{2M} F_2^q(0)=\frac{1}{2M} \kappa_{q/p}.
\ee
Here $e_q\kappa_{q/p}$ is the contribution from flavor $q$ to the
anomalous magnetic moment of the proton. Neglecting the contribution
from heavier quarks to the nucleon anomalous magnetic 
moment, one can use the proton and neutron anomalous
magnetic moment to solve for the contributions from $q=u,d$, yielding
$\kappa_{u/p} \approx 1.67$ and $\kappa_{d/p}\approx -2.03$. 
The resulting significant deformation ($|d_q|\sim 0.1 \mbox{fm}$)
of impact parameter dependent PDFs in the transverse direction 
(fig. \ref{fig:distort}), which is in opposite directions for
$u$ and $d$ quarks, should have observable consequences in other
experiments as well as will be discussed in the following. 

\section{Transverse Single-Spin Asymmetries}
In a DIS experiment on a 
transversely polarized target, the (on average attractive) final
state interactions (FSI) should cause a transverse momentum asymmetry
that is opposite to the transverse position space asymmetry.
This simple argument leads to the prediction \cite{mb:SSA} that the 
Sivers function $f_{1T}^{\perp q}(x,{k}_T^2)$
parameterizing the transverse momentum asymmetry
\cite{sivers,trento}
\be
f_{q/p^\uparrow}(x,{\bf k}_T) = f_1^q(x,k_T^2)
-f_{1T}^{\perp q}(x,k_T^2) \frac{ ({\bf {\hat P}}
\times {\bf k}_T)\cdot {\bf S}}{M},
\label{eq:sivers}
\ee
has the opposite sign as $\kappa_{q/p}$. This prediction was
confirmed in the {\sc Hermes} experiment \cite{hermes}. It is also consistent with
a vanishing Sivers function for a deuterium target \cite{compass}.

\section{Transverse Force on Quarks in DIS}

The polarized twist three parton distribution
$g_2(x)$ can be expressed as a sum of a piece that is
entirely determined in terms of $g_1(x)$ plus an interaction
dependent twist-3 part $\bar{g}_2(x)$ that involves quark gluon correlations
\cite{WW}
\bea
g_2(x)=g_2^{WW}(x)+\bar{g}_2(x) \label{eq:WW}\quad\quad\mbox{with}
\quad\quad
g_2^{WW}(x)=-g_1(x)+\int_x^1 \frac{dy}{y}g_1(y).
\eea
Here we have neglected $m_q$ for simplicity. For example,
its $x^2$ moment yields \cite{Shuryak,Jaffe}
\be
\int dx x^2 \bar{g}_2(x)= \frac{d_2}{3} \label{eq:d2}
\ee
with
\be
g\left\langle P,S \left|\bar{q}(0)G^{+y}(0)\gamma^+q(0) 
\right|P,S\right\rangle =
2 M {P^+}P^+ S^x d_2
\label{eq:twist3}.
\ee
In the limit where $Q^2$ is so low that the virtual photon 
wavelength is larger
than the nucleon size,  the electro-magnetic
field associated with the two virtual photons appearing
in the forward Compton amplitude corresponding to the structure 
function is nearly homogenous across the nucleon and the
spin-dependent structure function $g_2(x,Q^2)$ can be related to
spin-dependent polarizabilities.
In contradistinction, in the Bjorken limit, the matrix elements
describing the moments of $g_2(x,Q^2)$ are given by local
correlation functions, such as (\ref{eq:twist3}).
Nevertheless, because of the abovementioned low $Q^2$ interpretation
of $g_2$, the {\em local} matrix elements appearing in 
(\ref{eq:twist3})  
\be
\chi_E 2M^2 {\vec S} = \left\langle P,S\right|
q^\dagger {\vec \alpha} \times g {\vec E} q \left| P,S\right\rangle
\quad\quad\quad\quad\quad
\chi_B 2M^2 {\vec S} = \left\langle P,S\right|
q^\dagger g {\vec B} q \left| P,S\right\rangle,
\label{eq:chi}
\ee
where
\be
d_2 = \frac{1}{4}\left(\chi_E-2\chi_M\right),
\label{eq:d2chi}
\ee
(note that $\sqrt{2}G^{+y}=B^x-E^y$)
are sometimes called color electric and 
magnetic polarizabilities \cite{Ji}. In the following we will 
discuss why, at high $Q^2$, a better semi-classical
interpretation for these matrix elements is that of a `force'.

In electro-magnetism, the $\hat{y}$-component of the Lorentz 
force $F^y$ acting on a particle with charge $e$ moving, with 
the speed of light along the $-\hat{z}$ direction, reads
\be
F^y = e\left[ {\vec E} + {\vec v}\times {\vec B}\right]^y
=e \left(E^y - B^x\right) = -e\sqrt{2}F^{+y},
\ee
which involves the same linear combination of
Lorentz components that also appears in the 
gluon field strength tensor in (\ref{eq:twist3}).
Therefore (\ref{eq:twist3}) implies a relation between
$d_2$ and the color Lorentz force on a quark that moves
(in a DIS experiment) with ${\vec v}\approx (0,0,-1)$ \cite{force}
\bea
\label{eq:QS3}
F^y(0)\equiv - \frac{\sqrt{2}}{2P^+}
\left\langle P,S \right|\bar{q}(0) G^{+y}(0)
\gamma^+q(0) \left|P,S\right\rangle%\\
= -{\sqrt{2}} MP^+S^xd_2
= -{M^2}d_2,
%\nonumber
\eea
where the last equality holds only in the rest frame 
($p^+=\frac{1}{\sqrt{2}}M$) and for $S^x=1$,
can be interpreted as the averaged transverse
force acting on the active quark
in the instant right after it has been struck by the virtual photon.

For a nucleon polarized in the $+\hat{x}$
direction, the $\gamma^+$-distribution for $u$ ($d$) is
shifted towards the $\pm\hat{y}$ direction. This observation
suggests that
these quarks also `feel' a nonzero color-electric force pointing
on average in the $\mp\hat{y}$ direction, i.e. one would
expect that $d_2$ is positive (negative) for $u$ ($d$) quarks.
This would also be consistent with the observed signs of
the corresponding Sivers functions \cite{hermes,compass}.

A measurement of the $x^2$-moment $f_2$  of the 
twist-4 distribution $g_3(x)$ \cite{color}
allows determination of the expectation value of a different
linear combination of 
Lorentz/Dirac components of the quark-gluon correlator appearing
in (\ref{eq:twist3}) \cite{f2}
\be
f_2 M^2S^\mu = \frac{1}{2} \left\langle p,S\right|
\bar{q}g\tilde{G}^{\mu \nu}\gamma_\nu q\left|p,S\right\rangle . 
\ee
Using rotational invariance, to relate various Lorentz components
one thus finds a linear combination of the matrix elements of
electric and magnetic quark-gluon correlators (\ref{eq:chi})
\be
f_2=\chi_E-\chi_M,
\ee
that differs from that in (\ref{eq:d2chi}).
In combination with (\ref{eq:twist3}) this 
allows a decomposition of the force into electric and magnetic
components $F^y= F^y_E+F^y_M$.

A relation similar to (\ref{eq:QS3}) can be derived for the
$x^2$ moment of
interaction dependent twist-3 part $\bar{e}(x)$ of the 
scalar PDF $e(x)$. The average 
transverse force at $t=0$ (right after being struck) on a
quark with transversity in the $+\hat{x}$ direction reads
\be
F^y(0) = \frac{1}{2\sqrt{2}p^+} g\left\langle p\right| 
\bar{q} \sigma^{+y}G^{+y}q\left|
p\right\rangle = \frac{1}{\sqrt{2}}MP^+S^x e_2 = \frac{M^2}{2} e_2
\equiv \frac{M^2}{2}\int_0^1 \!\!dx \bar{e}(x)x^2
\label{eq:Fe2}
\ee
(in the rest frame of the target nucleon and for $S^x=1$).

The impact parameter distribution for quarks with transversity in
the $+\hat{x}$ direction was found to be shifted in the
$+\hat{y}$ direction \cite{DH,hagler,mb:brian}.
The chromodynamic lensing model \cite{mb:SSA} thus implies a force
in the negative $-\hat{y}$ direction for these quarks and one
thus expects $e_2<0$ for both $u$ and $d$ quarks. 
Furthermore, since 
$|\kappa_\perp|>|\kappa|$, one expects $|e_2| > |d_2|$.

\begin{theacknowledgments}
I would like to thank 
A.Bacchetta, D. Boer, S.J. Brodsky, J.P. Chen, Y.Koike, 
and Z.-E. Mezziani for useful discussions. 
This work was supported by the DOE (DE-FG03-95ER40965).
\end{theacknowledgments}

%%%%%%%%%%%%%%%%%%%%%%%%%%%%%%%%%%%%%%%%%%%%%%%%
%% The bibliography can be prepared using the BibTeX program or
%% manually.
%%
%% The code below assumes that BibTeX is used.  If the bibliography is
%% produced without BibTeX comment out the following lines and see the
%% aipguide.pdf for further information.
%%
%% For your convenience a manually coded example is appended
%% after the \end{document}
%%%%%%%%%%%%%%%%%%%%%%%%%%%%%%%%%%%%%%%%%%%%%%%%

\bibliographystyle{ws-procs9x6}
%\bibliography{ws-pro-sample}

\end{document}